\documentclass[twocolumn,showpacs,amsmath,amssymb]{revtex4}

\usepackage{graphicx}
\usepackage{dcolumn}
\usepackage{bm}

\begin{document}

\title{Asymptotic function for multi-growth surfaces using power-law noise}

\author{Hiroaki Katsuragi}
\email{katsurag@asem.kyushu-u.ac.jp}
\author{Haruo Honjo}

\affiliation{Department of Applied Science for Electronics and Materials, Interdisciplinary Graduate School of Engineering Science, Kyushu University, 6-1 Kasugakoen, Kasuga, Fukuoka 816-8580, Japan}

\date{\today}

\begin{abstract}
Numerical simulations are used to investigate the multiaffine exponent $\alpha_q$ and multi-growth exponent $\beta_q$ of ballistic deposition growth for noise obeying a power-law distribution. The simulated values of $\beta_q$ are compared with the asymptotic function $\beta_q = \frac{1}{q}$ that is approximated from the power-law behavior of the distribution of height differences over time. They are in good agreement for large $q$. The simulated $\alpha_q$ is found in the range $\frac{1}{q} \leq \alpha_q \leq \frac{2}{q+1}$. This implies that large rare events tend to break the KPZ universality scaling-law at higher order $q$.
\end{abstract}

\pacs{05.40.-a, 05.40.Ca, 05.40.Fb, 05.45.Df}

\maketitle

\section{Introduction}
Growing rough surfaces occur everywhere in nature and are encountered in engineering and everyday life. Examples include, fluid displacement in porous media \cite{Rubio01}, the growth of crystals \cite{Honjo01} or colonies of bacteria \cite{Ben-Jacob01}, the propagation of a wet front on paper \cite{Buldyrev01}, and so on \cite{Barabasi01}. A growing rough surface is one of the simplest patterns created by a nonequilibrium state. Consequently, many studies have examined growing rough surfaces. Two exponents characterize a growing rough surface. One is the roughness exponent $\alpha$, which relates the space length scale $x$ to the surface width $w$, ($w \sim x^\alpha$). The other is the growth exponent $\beta$, which relates the time scale $t$ to the surface width, ($w \sim t^\beta$). A scaling function including these two exponents can be written as $w \sim x^\alpha \Psi(\frac{t}{x^z})$, where $z=\frac{\alpha}{\beta}$ is the dynamic exponent \cite{Family01}. There are two different scaling regimes in $\Psi$ depending on the argument $u=\frac{t}{x^z}$. $\Psi(u) \sim u^\beta$ when $u \ll 1$, and $\Psi(u) \sim \mbox{const.}$ when $u \gg 1$.

The KPZ (Kardar-Parisi-Zhang) equation was proposed as an equation that models the dynamics of a growing rough surface \cite{Kardar01}. The KPZ equation is written as $\partial_t h = \sigma \nabla^2 h + \frac{\lambda}{2} (\nabla h)^2 + \eta$, where $h$, $\sigma$, $\lambda$, and $\eta $ are the surface height, effective surface tension, lateral growth strength, and a noise term, respectively. Since the KPZ equation is a stochastic nonlinear differential equation, an exact solution cannot be obtained. However, the roughness and growth exponents can be calculated by the renormalization group method as $\alpha=\frac{1}{2}$ and $\beta=\frac{1}{3}$, respectively \cite{Kardar01}. It has been suggested that the scaling-law $\alpha + z = 2$ holds in the KPZ universality class. Although most numerical simulations follow this result, many experiments of growing rough surfaces show $\alpha \simeq 0.75 \sim 0.85 > \frac{1}{2}$ \cite{Barabasi01}. However, the KPZ universality scaling law is satisfied even in many experiments whose $\alpha$ is larger than $\frac{1}{2}$. In order to interpret the large $\alpha$, several models have been proposed. These include the power-law noise model \cite{Zhang01}, quenched noise model \cite{Csahok01}, and correlated noise model \cite{Lam01}. Moreover, certain systems with a large $\alpha$ obey EW (Edwards-Wilkinson) universality \cite{Edwards01}. In this paper, we focus on the power-law noise model.

The noise in KPZ growth is ordinarily considered uncorrelated Gaussian noise. Zhang suggested KPZ growth with uncorrelated power-law noise \cite{Zhang01}. He performed numerical simulations of a BD (ballistic deposition) model and found that $\alpha$ varies with the exponent of the power-law noise $\mu$.  Power-law behavior of the noise distribution was observed in a fluid flow experiment \cite{Horvath02}. Therefore, the Zhang model is considered a moderate model for growing rough surfaces.

On the other hand, Barab\'asi et al.\ investigated multiaffinity of the BD model with power-law noise \cite{Barabasi02}. Multiaffine analysis is defined by the $q$th order height-height correlation function $C_q(x)$ as $C_q (x) = \langle |h(x')-h(x'+x)|^q \rangle_{x'} \sim x^{q\alpha_q}$, where $\alpha_q$ is the $q$th order roughness exponent.

While the entropy spectrum method describes the {\em dynamic} characteristics of {\em self-similar} fractals \cite{Honjo02}, it describes the {\em static} characteristics of {\em self-affine} fractals \cite{Katsuragi01}. That is, multiaffine analysis is equivalent to the entropy spectrum method for self-affine fractals \cite{Katsuragi01}. Therefore, we must discuss the behavior of the $q$th order growth exponent $\beta_q$ in order to examine the dynamic characteristics of growing self-affine fractals. In ordinary variance analysis for the case $q=2$, a mean field approximation has been applied to obtain the relations among $\alpha$, $\beta$, and $\mu$ \cite{Zhang02}, and the scaling-law $\alpha + z = 2$ was confirmed.

Myllys et al.\ investigated the slow combustion of paper and reported that $\alpha_q$ and $\beta_q$ at the combustion front varied with $q$ \cite{Myllys01}. Based on their discussion, we can expand the $q$th order height-height correlation function as
\begin{equation}
C_q (x,t) = \langle |\delta h(x',t') - \delta h(x'+x,t'+t)|^q \rangle_{x',t'}, \label{eq:Cq(x,t)def}
\end{equation}
where $\delta h(x,t) \equiv h(x,t) - \langle h(t) \rangle_x$. Then, we can define $\alpha_q$ and $\beta_q$ as
\begin{subequations}
\begin{equation}
C_q(x,0) \sim x^{q \alpha_q}, \label{eq:alphaqdef}
\end{equation}
\begin{equation}
C_q(0,t) \sim t^{q \beta_q}. \label{eq:betaqdef}
\end{equation}
\end{subequations}
The exponents $\alpha_q$ and $\beta_q$ are defined at the limits $x \to 0$ and $t \to 0$, respectively.

The above-mentioned numerical simulations did not examine the $q$ dependences of $\alpha_q$ and $\beta_q$ on various $\mu$ thoroughly. Barab\'asi studied the temporal fluctuation of the surface width, however, his analysis examined the period after the saturation of surface growth \cite{Barabasi04}. Therefore, we investigate the behavior of $\alpha_q$ and $\beta_q$ directly to determine how $\alpha_q$ and $\beta_q$ depend on the noise parameter $\mu$ for higher order $q$. We are also interested in whether the KPZ universality scaling law $\alpha+z=2$ is still valid for higher order $q$. Here, we report the results of numerical simulations of the BD model, and compare simulated data and the calculated asymptotic functions for $\alpha_q$ and $\beta_q$.

\section{Simulations}
We investigate a $1+1$ dimensional BD model whose growing dynamics are described by
\begin{eqnarray}
h(x,t+1) &=& \max[h(x-1,t)+\eta(x-1,t), h(x,t) \nonumber \\
&& +\eta(x,t),h(x+1,t)+\eta(x+1,t)]. \label{eq:BDrule}
\end{eqnarray}
This ultra-discrete BD algorithm can be connected to the KPZ equation \cite{Nagatani01}. We start with $h(x,0)=0$ for all $x$, and the surface evolves according to Eq.\ \ref{eq:BDrule}.  We use periodic boundary conditions for the space dimension. The uncorrelated noise $\eta$ is taken from a power-law distribution in the form \cite{Zhang01}
\begin{equation}
P(\eta) \sim \frac{1}{\eta^{(\mu+1)}} \; \mbox{ for } \eta>1; \quad P(\eta)=0 \; \mbox{ otherwise}. \nonumber
\end{equation}

We focus on the range $2 \leq \mu \leq 5$. The variance of $P(\eta)$ is finite for $\mu > 2$ and its statistical properties differ from those of Gaussian noise for $\mu \leq 5$ \cite{Zhang01, Zhang02}.

\begin{figure}
\scalebox{1.0}[1.0]{\includegraphics{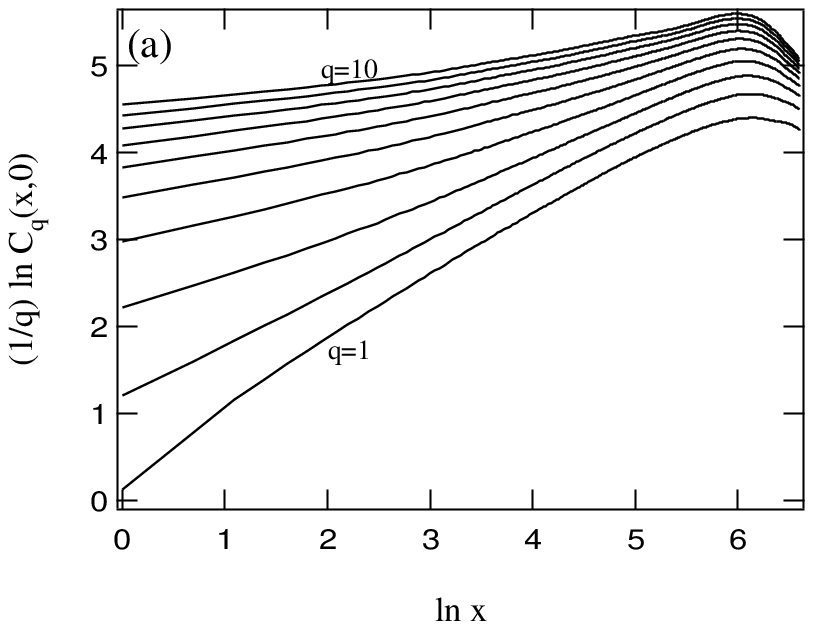}}
\scalebox{1.0}[1.0]{\includegraphics{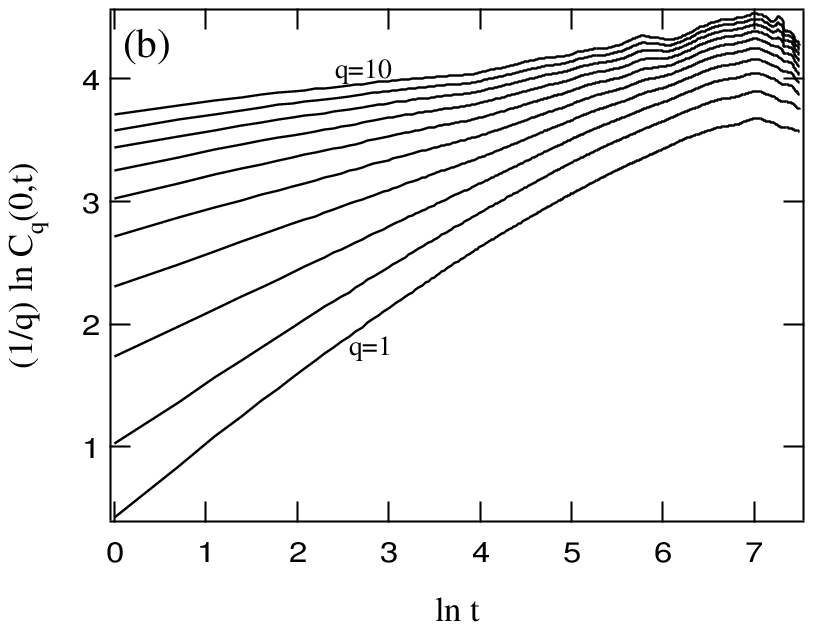}}
\caption{
The $q$th order height-height correlation functions at $\mu=3.0$. (a) $\frac{1}{q} \ln C_q(x,0)$ vs. $\ln x$. (b) $\frac{1}{q} \ln C_q(0,t) $ vs. $\ln t$. In the figures, the curves correspond to $q=1,2, \ldots, 10$, from bottom to top.
}
\label{Cq}
\end{figure}

\begin{figure}
\scalebox{1.0}[1.0]{\includegraphics{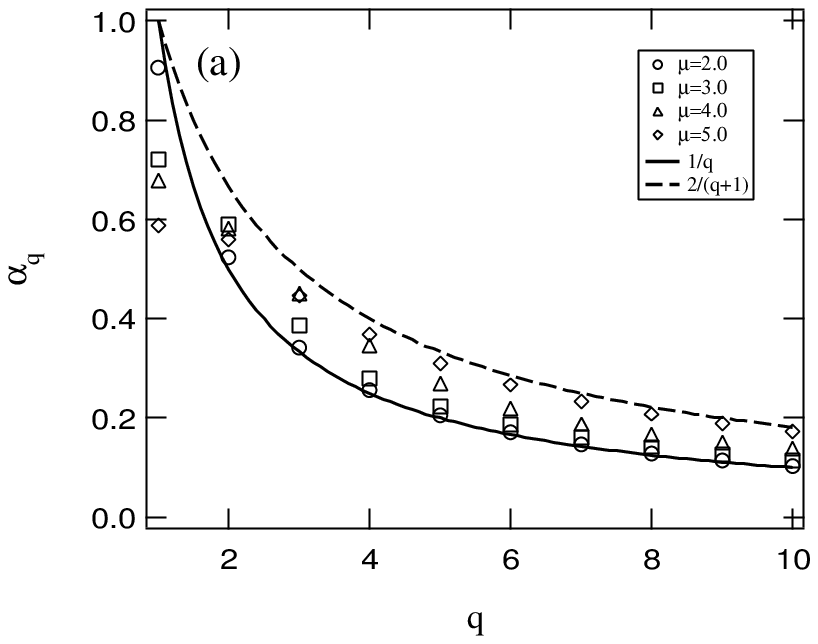}}
\scalebox{1.0}[1.0]{\includegraphics{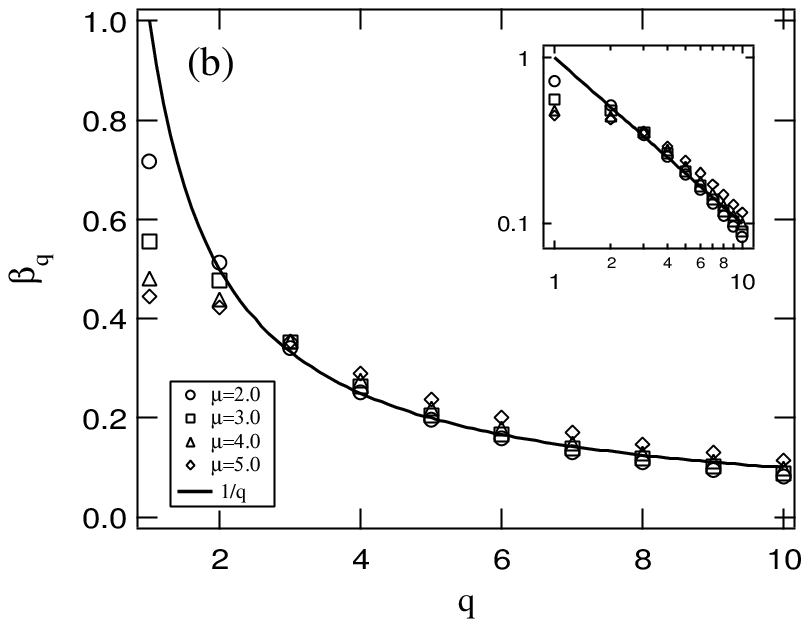}}
\caption{
The q dependences of (a)$\alpha_q$ and (b)$\beta_q$. The solid and broken lines correspond to the asymptotic functions $\frac{1}{q}$ and $\frac{2}{q+1}$, respectively. The inset in (b) is a log-log plot.
}
\label{alphabeta}
\end{figure}

First, we carry out a BD model simulation to investigate the $q$th order height-height correlation functions. Figure \ref{Cq} shows the behavior of $C_q(x,0)$ and $C_q(0,t)$ at $\mu=3.0$.  From bottom to top, the curves correspond to $q=1, 2, \ldots, 10$. In Fig.\ \ref{Cq}, we use the system size $L=1024$, the time step $T=$(a)$5000$, (b)$2000$, and the average ensemble number $N=$(a)$100$, (b)$5$. The exponents $\alpha_q$ and $\beta_q$ are obtained from the slopes of the curves in Fig.\ \ref{Cq}. The values of $\alpha_q$ and $\beta_q$ are defined as the limit of $x \to 0$ and $t \to 0$, respectively (Eqs.\ \ref{eq:alphaqdef} and \ref{eq:betaqdef}). In addition, for regions of large $t$ and $x$, conventional non-multiaffine scaling appears \cite{Barabasi02}. Therefore, the fitting regions are approximated as $\ln t \leq 3$ and $\ln x \leq 3$. Fig.\ \ref{alphabeta} shows the $q$ dependences of $\alpha_q$ and $\beta_q$ on various $\mu$. For large $q$ in Fig.\ \ref{alphabeta}(b), $\beta_q$ seems to collapse independently of $\mu$. Since most of the error for each point is within the symbol used, there are no error bars in Fig.\ \ref{alphabeta}.

When we use noise with a Gaussian distribution, we obtain the fully parallel curves in plot Fig.\ \ref{Cq}. This indicates that Gaussian noise and power-law noise affect the $q$th order moment analysis differently, and that the system size is sufficient to calculate the $q$th order moment.

\begin{figure}
\scalebox{1.0}[1.0]{\includegraphics{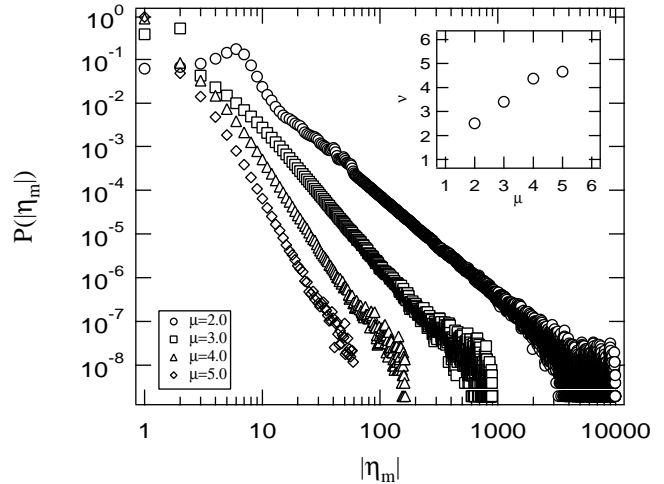}}
\caption{
The distribution of height differences between neighbour times. The relation between $\mu$ (input noise parameter) and $\nu$ (measured noise parameter) is plotted in the inset.
}
\label{heightdiff}
\end{figure}

Next, we measure the amplitude of the effective noise $\eta_m \equiv \delta h(x,t+1) - \delta h(x,t)$ from the growth patterns directly \cite{Myllys01}. Fig.\ \ref{heightdiff} shows a log-log plot of the probability distribution $P(|\eta_m|)$ vs. $|\eta_m|$. The parameters used are $L=1024$, $T=5000$, and $N=100$. $P(|\eta_m|)$ clearly shows power-law behavior, but these exponents are slightly different from $-(\mu+1)$. Since these distributions obey the power-law, we denote the exponent as $-\nu$ in this paper. This difference between $-(\mu+1)$ and $-\nu$ might result from the lateral growth effect of the BD model, which corresponds to the nonlinear term in the KPZ equation.

\section{Asymptotic function}
\begin{figure}
\scalebox{0.7}[0.7]{\includegraphics{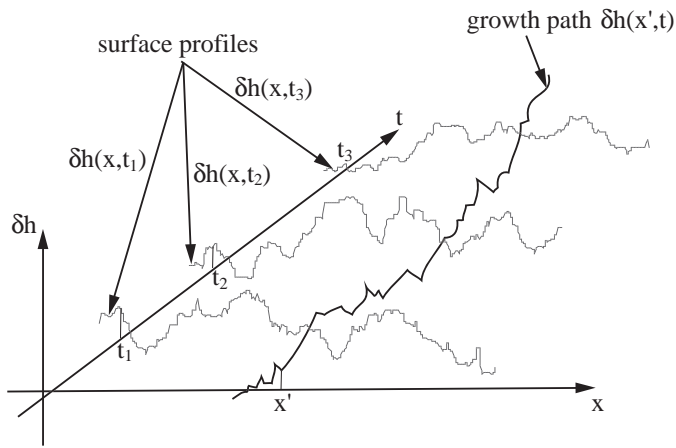}}
\caption{
A schematic image of the growth path. The surface profile $\delta h(x,t')$ corresponds to a snapshot of the surface at some time $t'$. The intersection between the rough surface $\delta h(x,t)$ and some position $x=x'$ is the growth path.
}
\label{gpath}
\end{figure}

Here, we calculate the multi-growth height-height correlation function $C_q(0,t)$ based on the multiaffine concept \cite{Barabasi03}. We discuss the growth path at some position $x'$: $\delta h(x',t)$. Fig.\ \ref{gpath} shows a schematic image of the growth path. The growth path $\delta h(x', t)$ is the intersection with the $\delta h(x,t)$ surface at some position $x=x'$.  We can generally normalize the time range and height to $T$ and $\delta h_{max} - \delta h_{min}$, respectively, i.e., we consider the path $\delta h(x',t)$ for the range $0 \leq t \leq 1$ and $0 \leq \delta h \leq 1$. We can use the local growth exponent $\gamma$ to characterize the local singularity of the surface growth $|\delta h(x',t'+\tau) - \delta h(x',t')| \sim \tau^{\gamma}$, where $\tau=T^{-1}$ and $\gamma \geq 0$. The number of height difference segments of length $l=|\delta h(x',t'+\tau) - \delta h(x',t')|$ in the growth path can be written as $l^{-\nu}$ from the results of Fig.\ \ref{heightdiff}. Therefore, the number of segments on growth path $N(\gamma)d\gamma$ that have singularity exponents in the range $(\gamma, \gamma + d\gamma)$ is found to scale with $\tau$ as $N_\tau(\gamma) \sim \tau^{-\gamma \nu}$. We can obtain the height-height correlation function for the limit $\tau \to 0$ as follows
\begin{eqnarray}
C_q(0,\tau) &\sim& \frac{1}{T} \int (\tau^{\gamma})^qN_\tau(\gamma) \rho(\gamma)d\gamma \nonumber\\
&\sim& \int \tau^{1 + \gamma q - \gamma \nu} \rho(\gamma)d\gamma. \label{eq:Cqcalc}
\end{eqnarray}
Where, the function $\rho(\gamma)$ is a density function independent of $\tau$. For a continuous system, the integral in Eq.\ \ref{eq:Cqcalc} must be dominated by the value of $\gamma$ that minimizes $1 + \gamma q - \gamma \nu$. Therefore, we replace $\gamma$ with the value $\gamma(q)$ and compare the exponent with Eq.\ \ref{eq:betaqdef},
\begin{equation}
\beta_q = \frac{1}{q} + \gamma(q) \bigl( 1 - \frac{\nu}{q} \bigr). \label{eq:betaqform}
\end{equation}

Note that $\gamma (q)$ decreases monotonically with increasing $q$, and $\gamma (q) \geq 0$ \cite{Feder01}. In the limiting case $q\gg1$, we assume that $\gamma (q)$ vanishes faster than $\frac{1}{q}$. Finally, we obtain a simple asymptotic form of $\beta_q$ as
\begin{equation}
\beta_q = \frac{1}{q}, \mbox{ at } q \gg 1. \label{eq:betaqasymp}
\end{equation}
We plot this function in Fig.\ \ref{alphabeta} as solid lines. While this function does not include any fitting parameter, the simulated data agree with this function for large $q$ in Fig.\ \ref{alphabeta}(b). If we assume that the relation $\alpha_q + \frac{\alpha_q}{\beta_q} = 2$ holds for higher order $q$, the asymptotic form of $\alpha_q$ can be calculated as:
\begin{equation}
\alpha_q = \frac{2}{q+1}, \mbox{ at } q \gg 1. \label{eq:alphaqasymp}
\end{equation}
This function is plotted in Fig.\ \ref{alphabeta}(a) as a broken line. The simulated values of $\alpha_q$ seem to distribute in the region $\frac{1}{q} \leq \alpha_q \leq \frac{2}{q+1}$ at large $q$.

\section{discussion}
The contribution of large segments in the integral of Eq.\ \ref{eq:Cqcalc} becomes dominant for higher order $q$. Since large segments are characterized by small $\gamma$ in our notation, $\gamma (q)$ decays rapidly due to the presence of large rare events with power-law distributed noise. Then, the effect of $\mu$ (or $\nu$) becomes negligible, as written in Eq.\ \ref{eq:betaqasymp}.  Moreover, Eq.\ \ref{eq:betaqform} becomes equivalent to Eq.\ \ref{eq:betaqasymp} at $q \simeq \nu$. Since the absolute value of the second term in Eq.\ \ref{eq:betaqform} is not small, $\beta_q$ deviates from Eq.\ \ref{eq:betaqasymp} at $q < \nu$. This qualitative change around $q \simeq \nu$ corresponds to the phase transition point of Barab\'asi et al.\ \cite{Barabasi02}. The term $\frac{1}{q}$ originates solely from the normalization factor $\frac{1}{T}$ in Eq.\ \ref{eq:Cqcalc}. The presence of large rare events is necessary, but the value of the exponent of noise distribution is not important for obtaining the asymptotic result Eq.\ \ref{eq:betaqasymp}.

In the early growth stage, we can confirm the KPZ universality scaling-law perfectly. For instance, we calculated $\alpha_q$ at $T=2000$ and found that all $\alpha_q$ approach the curve for Eq.\ \ref{eq:alphaqasymp} independent of $\mu$. In addition, a crossover to conventional non-multiaffine scaling at large $x$ is clearly observed. We can also observe power-law-like tails of the distribution of the height difference $\delta h(x+1,t) - \delta h(x,t)$. Then, $\alpha_q$ leads to the function $\frac{1}{q}$ for the limit $T \to \infty$ using the same calculation as for $\beta_q$. Since the non-uniformity of the power-law noise increases as $\mu$ decreases, the influence of large rare events dominates in the small $\mu$ system. Therefore, the smaller $\mu$ becomes, the nearer $\alpha_q$ approaches $\frac{1}{q}$.  Namely, $\alpha_q$ obeys the KPZ universality scaling law $\frac{2}{q+1}$ in the early growing stage, and obeys the rare event dominant behavior $\frac{1}{q}$ in the fully developed stage.

In the inset of Fig.\ \ref{alphabeta}(b), a slight discrepancy between the simulated data and the function $\frac{1}{q}$ is observed. The discrepancy is due to the small, but finite, $\gamma(q)$ effect. Since the probability of finding a large event decreases as $\mu$ increases, the value $\gamma(q)$ does not decay quickly when $\mu$ is large. Therefore, the statistical property for large $\mu$ deviates from the power-law noise case and approaches the Gaussian noise case \cite{Zhang02}. The range $2 \leq \mu \leq 5$ is considered the power-law dominant range. In this sense, Eq.\ \ref{eq:betaqasymp} might be a limited approximation form. The precise correction to use for the range of small $\mu$ (in particular, the L\'evy distribution case $\mu \leq 2$) remains unsolved.

In a paper combustion front experiment, $\beta_q$ varied across the value $0.5$, and $\alpha_q$ did not fall beneath $0.5$ \cite{Myllys01}. This tendency in $\beta_q$ is similar to our result. Myllys et al.\ found the effective power-law noise amplitude in their system to be $3.72 \leq \nu \leq 5.0$. Our simulations include this range. Then, the behavior of $\beta_q$ inevitably approaches the form of Eq.\ \ref{eq:betaqasymp}.  However, the behavior of $\alpha_q$ differs from our result. This means that the paper combustion front grows according to power-law noise, but breaks the KPZ universality in a different manner from our result. This experimental system might belong to another universality class. In other experimental systems, $q$th order exponents should be measured in order to discuss universality classification in detail. There have been no multiaffine analyses of other models, such as the quenched noise, correlated noise, or EW class models. Both numerical and experimental studies are needed to further understand growing rough surfaces.

In summary, we performed numerical simulations of the BD model with power-law distributed noise $(2 \leq \mu \leq 5)$. The simulated $q$th order growth exponent is in good agreement with the approximate asymptotic function $\beta_q = \frac{1}{q}$ for higher order $q$.  Assuming the KPZ universality scaling-law $\alpha_q + \frac{\alpha_q}{\beta_q} = 2$, we obtain $\alpha_q = \frac{2}{q+1}$. $\alpha_q$ is distributed in the region $\frac{1}{q} \leq \alpha_q \leq \frac{2}{q+1}$. This indicates that $\alpha_q$ is determined by competition between large rare events and KPZ universality.

\end{document}